\documentclass{elsart}
\usepackage{epsfig}
\usepackage{lineno}

\def\co2{CO$_2$}
\journal{Nuclear Instruments and Methods}
\begin{document}
\newcommand{\cstwo}{CS$_2\,\,$  }
\begin{frontmatter}

\title{A benign, low Z electron capture agent for negative ion TPCs}
\author[temple]{C.J. Martoff\corauthref{cor}},
\corauth[cor]{Corresponding author.}
\ead{jeff.martoff@temple.edu}
\ead{(215)204-3180}
\author[temple]{M.P. Dion},
\author[temple]{M. Hosack\thanksref{now}},
\thanks[now]{Present address:Bruker AXS Inc.,
5465 East Cheryl Parkway Madison, WI  53711, USA
}
\author[temple]{D. Barton}

\address[temple]{Department of Physics, Temple Universiy, Philadelphia, PA 19122, USA}
\author[Black]{J.K. Black}
\address[Black]{Rock Creek Scientific, 1400
East{}-West Hwy, Suite 807, Silver Spring, MD 20910 USA}

\begin{keyword}
\sep Dark Matter \sep negative ion\sep X-ray polarimeter \sep diffusion \sep ion mobility 
\PACS 29.40.Cs \sep 51.20.+d \sep 51.50.+v \sep 52.70.La \sep 95.35.+d \sep 95.55.Qf
\end{keyword}

\begin{abstract} 
We have identified nitromethane (CH$_3$NO$_2$)
as an effective electron capture agent for negative ion TPCs (NITPCs).
\ We present drift velocity and longitudinal diffusion measurements for
negative ion gas mixtures using nitromethane as the capture agent.
\ Not only is nitromethane substantially more benign than the only
other identified capture agent, CS$_2$, but its low atomic
number will enable the use of the NITPC as a photoelectric X{}-ray
polarimeter in the 1{}-10 keV band.
\end{abstract}

\end{frontmatter}
\section{Introduction} The negative ion time projection chamber (NITPC) achieves high spatial
resolution by transporting charge in the form of negative ions, rather
than electrons, thereby reducing diffusion to the thermal limit in both
the transverse and longitudinal drift directions\cite{orignitpc}. 
\ 
This provides
the highest 3{}-D space{}-point resolution attainable for long drift
distances, particularly where operation at low pressures is required to
make very low energy tracks long enough to measure (for example, dark
matter WIMP recoil atoms). \ In relatively low rate environments, the
NITPC also has the advantage that the ion drift velocities are about
two orders{}-of{}-magnitude less than those of electrons. \ This allows
the use of low{}-bandwidth electronics that results in lower noise and
power consumption.

The key to reducing diffusion to the thermodynamic lower limit (Equation
1 below) is to prevent net energy transfer from the drift field to the
drifting primary ionization\cite{blumrolandi}. \ Such ``heating'' 
of the drifting
charge occurs when electrons are drifted at reduced fields E/p
{\textgreater} $\sim${}10{}-50 V/m${\cdot}$Torr. \ Heating is reflected
in an increase of the parameter ${\epsilon}$ in Equation 1 to values
higher than $\frac{3}{2}k$T, resulting in increased diffusion. \ For electron
drift, gas additives with large integrated cross sections for inelastic
scattering of electrons, such as hydrocarbons and CO$_2$,
provide channels for dissipation and are known \cite{blumrolandi,saulibible} 
to prevent heating of
drifting electrons up to moderate values of the reduced field ($\sim${}50
V/m${\cdot}$Torr). 

In contrast, ions drifting in even a high field are very efficiently
thermally coupled to the room{}-temperature gas, since the ion mass is
comparable to that of the gas molecules. \ Then, even elastic
scattering produces substantial energy transfer from the drifting ion
to the bulk gas. \ Ion diffusion has been measured in a number of
CS$_2$ mixtures suitable for TPC 
operation\cite{orignitpc,onebar,ohnukics2} and 
was
found to follow the ``thermal, low field" limiting
behavior\cite{blumrolandi}:

\begin{equation}
  \sigma _{D}=\sqrt{\frac{4\epsilon L}{3eE}}   
\label{bkm:BMEqndiff}
\end{equation}

Here $\sigma_D$ is the rms diffusion spread for carriers with charge $e$
after drifting through a distance $L$ in electric field $E$ with 
average (thermal) energy $\epsilon$.
The parameter \textit{${\epsilon}$} (the average thermal energy of a
drifting charge carrier) remained approximately equal to the
room{}-temperature value , up to reduced drift fields as high as $\sim${}2.5 $\times$ 10$^{3}$
V/m${\cdot}$Torr.

Many chemical vapors at STP will capture thermal electrons to form
negative ions\cite{chembook}. \ However, to maintain good track 
resolution the
agent must have a capture mean free path of a hundred microns or less
at a relatively low partial pressure. \ Since gas gain is desired at
the TPC endcaps, the agent must also relinquish its electrons in the
high, strongly inhomogeneous field of the gain structures, allowing
Townsend avalanches to form. \ In all previous work,
CS$_2$ has been used as the only capture agent fulfilling
these requirements reasonably well.

While the NITPC has been successfully implemented with
CS$_2$ as the capture agent in dark matter 
observatories\cite{drift1,drift2,drift3}, the utility of the NITPC would 
be expanded by the identification of new capture agents. \ CS$_2$ has a combination of low flash 
point (-30$^{\circ}$ C),
high vapor pressure (400 Torr) and low explosive mixture limit in air (1.3\%)which
require great care in handling the material\cite{msdscs2}. Nitromethane\cite{msdsnitro}
has a flash point far above room temperature and a vapor pressure 
less than a tenth that of CS$_2$.  The comparative health effects of the two 
materials are indicated by the respective ACGIH Threshold Limit 
Values of 1 ppm for CS$_2$ vs. 20 ppm for nitromethane. 
CS$_2$ also may form sulfide deposits under conditions of high rate or discharge. \ More importantly, new applications for the NITPC would be enabled by capture agents with
different functional properties. \ For example, a single{}-ion counting
NITPC with high energy resolution for double beta decay searches has
been proposed\cite{Nygren07} that would benefit from a capture agent with higher ion
mobility than CS$_2$. \ Another example is photoelectric X{}-ray
polarimetry below 10 keV with a NITPC, which requires capture agents with
lower atomic number (see below).

Photoelectric polarimeters derive their information by determining the
photoelectron emission direction, which is strongly correlated with the
electric field vector of the incident photon. \ The most sensitive
photoelectric polarimeters image the tracks of the photoelectrons with
pixels small compared to the photoelectron track length and reconstruct
the emission direction on an event{}-by{}-event basis. \ Such
polarimeters have been realized with optical avalanche 
chambers\cite{pol1,pol2},
micropattern gas detectors with direct pixel 
readout\cite{pol3,pol4,pol5}, and a micropattern TPC\cite{Black07a}.

X{}-ray polarimetry below 10 keV with a NITPC would benefit from low atomic number
capture agents for several reasons. \ First, the correlation between the
photoelectron emission direction and photon electric field is
strongest for emission from s{}-orbitals\cite{Cooper}, so it is preferable to use
gases whose K{}-shell energies are well below the X{}-ray energy.
\ Second, the photoelectron is emitted with a kinetic energy that is
the difference between the binding energy and the X{}-ray energy so
that \ lower Z gases result in photoelectrons with larger kinetic
energies and longer track lengths.  Finally, lower Z gases have less multiple Coulomb scattering, which results in straighter tracks and a better correlation between the
photoelectron emission direction and photon electric field.

Therefore, motivated by the requirements of low energy X{}-ray
polarimetry and guided by the chemical literature on electron capture,
we have explored a number of compounds as alternative NITPC capture
agents. \ We have identified nitromethane
(CH$_3$NO$_2$) as a suitable capture agent
for X{}-ray polarimetry which is also relatively benign. \ We present
measurements of drift velocity, gas gain and longitudinal diffusion in gases with
nitromethane as the capture agent.

\section{Capture agent screening apparatus}
Capture agents were screened for negative ion drift velocity, gas gain
and longitudinal diffusion at low to moderate fields using a
single{}-wire proportional counter attached to a homogeneous{}-field
drift region. \ The apparatus was enclosed in a stainless steel bell
jar with a simple gas manifold. Drift fields up to 4.0 $\times$ 10$^{4}$ V/m were used
with this apparatus at reduced pressure. \ Photoelectrons were generated 
at an Sn photocathode attached to the
drift{}-cathode. \ The photocathode was arranged so that it could be
cleaned between runs using a glow discharge in pure argon. \ This was
essential for maintaining the photoelectron yield.

UV light flashes from an EG\&G Flash{}-Pak\cite{egg} were admitted 
into the bell
jar through a quartz window, passed through a hole in the proportional
counter wall, and struck the photocathode producing photoelectrons.
\ The standard internal capacitors of the Flash{}-Pak were augmented
with additional HV capacitors to give a stored energy of about 0.2
Joule per pulse. \ The Flash{}-Pak was triggered by an external pulser,
from which a time{}-zero signal was also derived. \ The proportional
wire signal was read out through an ORTEC 142PC preamp and an ORTEC 579
shaping amplifier. \ Drift times to the proportional wire and
wire{}-signal time widths were measured with a digital oscilloscope.

Drift velocities and gas gain at moderate to high fields (1.0 $\times$ 10$^{5}$ {}- 4.0 $\times$ 10$^{5}$ V/m) were
measured in various gas mixtures with a micropattern gas detector
(MPGD) having the same electrode structure as the gas electron
multiplier\cite{SauliGEM}, but assembled from two etched stainless 
steel meshes\cite{Black07a}
separated by a 100 ${\mu}$m thick teflon spacer. \ A drift
electrode, also of stainless steel, was placed 10 mm above the MPGD
cathode. \ The MPGD was operated at a gain of about 3000. \ The MPGD
cathode was instrumented with a charge{}-sensitive preamplifier
followed by a bipolar shaping amplifier with a six microsecond shaping
time constant.

The MPGD cathode and drift electrode were simultaneously illuminated
with a xenon flashlamp (Perkin{}-Elmer PAX{}-10), which produced
photoelectrons from both surfaces. \ Photoelectrons from the cathode
produced a prompt pulse, while the photoelectrons from the drift
electrode produced negative ions and gave a pulse delayed by the 
drift time across the 10mm
drift distance. \ The drift time was measured by averaging 100 pulses
on a digital oscilloscope and taking the difference between the
zero{}-crossing times of the prompt and the delayed pulses.

\section{Results and Discussion}
Nitromethane is known to have a large capture cross section for thermal
electrons\cite{chembook}. \ However, at moderate to high drift fields the pure
near{}-saturated vapor at 20 Torr produces both  electron{}-drift and
negative{}-ion{}-drift signals. \ This is understandable since a drift
field of  5.0 $\times$ 10$^{4}$ V/m at 20 Torr is already a reduced 
field of 2500 V/m$\cdot$Torr, easily high enough to raise the energy of drifting electrons out
of the thermal range before they could be captured. CO$_2$ 
was therefore
added to the nitromethane vapor. CO$_2$ is known to be
very effective in thermalizing drifting electrons, due to its large
inelastic scattering cross sections\cite{saulibible}. Mixtures of 20 Torr nitromethane
with 50 Torr or more of CO$_2$ were found to have
satisfactory characteristics as negative ion drift mixtures. 

\begin{figure}[h]
\begin{center}\thicklines
\epsfig{file=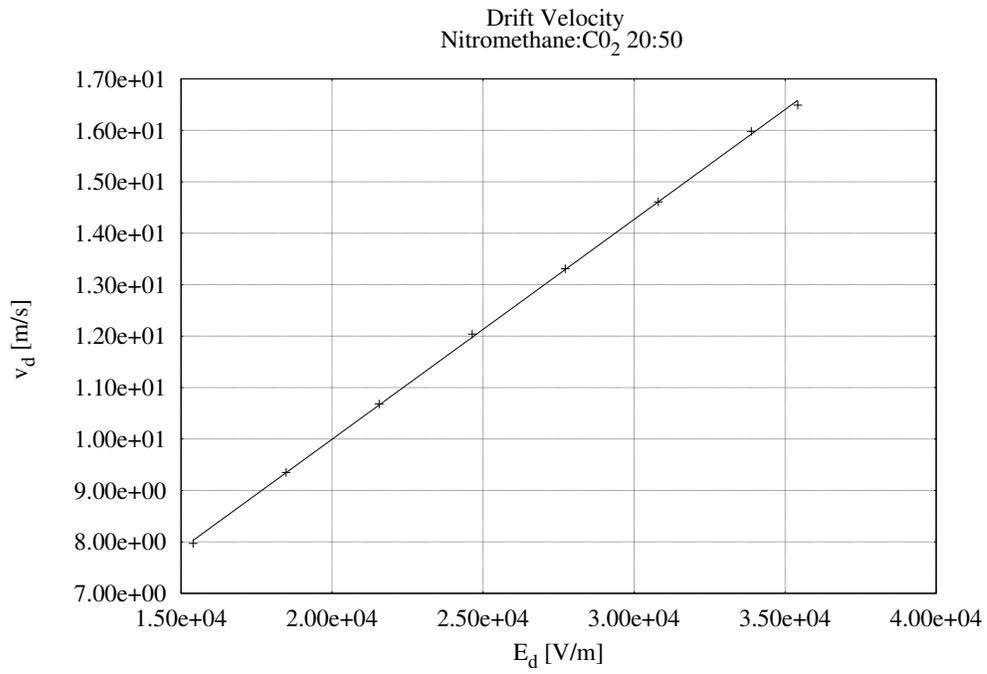,height=3.7in}
\caption{Drift velocity $v_d$ vs. drift field $E_d$ for 
Nitromethane:\co2 20:50}
\label{Fig:low_vel}
\end{center}
\end{figure}

\begin{figure}[h]
\begin{center}\thicklines
\epsfig{file=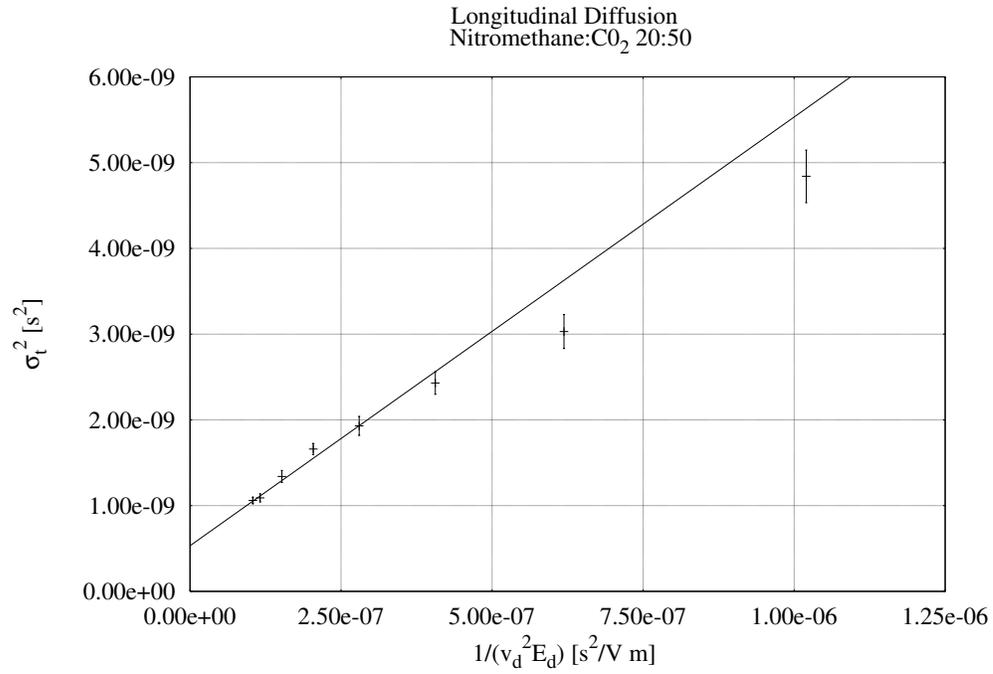,height=3.7in}
\caption{Longitudinal diffusion data for Nitromethane:\co2 20:50, 
plotted as
the square of the pulse time width vs. 1/$v_d^2 E_d$ for 80 mm 
drift.}
\label{Fig:diff}
\end{center}
\end{figure}

Drift velocity and diffusion results for the 
nitromethane:CO$_2$ 20:50
mixture are shown in Figures \ref{Fig:low_vel}  and 
\ref{Fig:diff}. \ The drift 
velocity rises linearly
with drift field, with mobility (4.27 ${\pm}$ .03) ${\times}$ 10$^{-4}$
m$^2$/V${\cdot}$sec and a small positive intercept (1.44 ${\pm}$ .08 m/s). \ The
diffusion is studied by plotting the square of the pulse width in time
as a function of 1/($v_d^2 E_d$). \ The pulse width is expected to be the
quadrature sum of the amplifier shaping time (10 ${\mu}$sec) and the
broadening due to diffusion. \ If the diffusion is governed by Equation
1, this plot will be linear in the region where diffusion dominates
(essentially the entire plot), with slope. \ Fitting 
Figure \ref{Fig:diff}  to a
straight line gives T = 313 ${\pm}$ 25 K.

\begin{figure}[h]
\begin{center}\thicklines
\epsfig{file=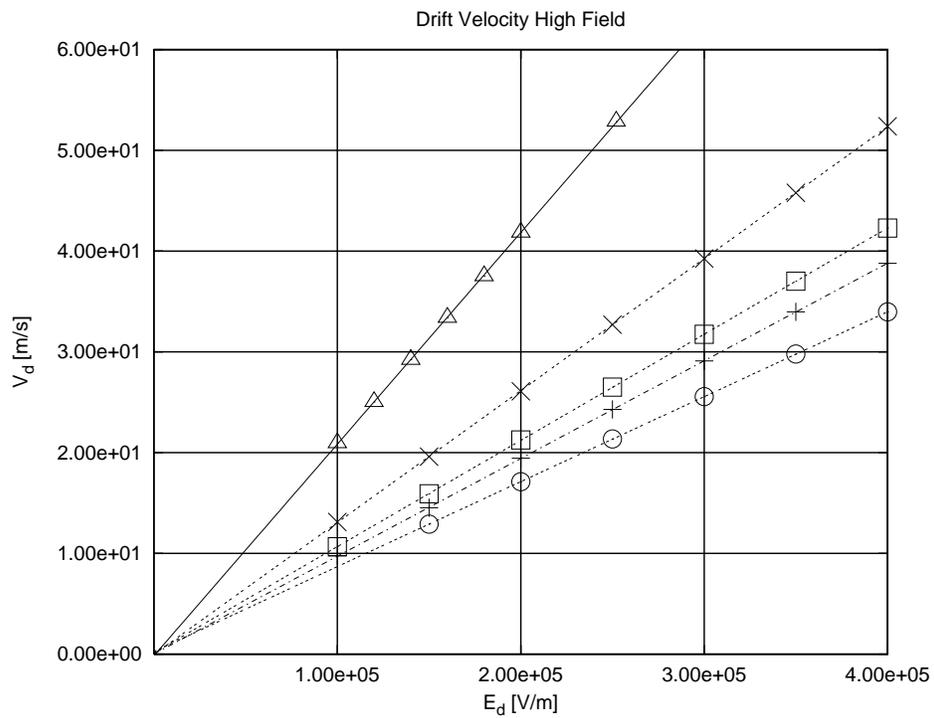,height=3.8in}
\caption{Drift velocity at high fields in mixtures of Nitromethane 
and
CO$_2$ and/or Neon, Methane mixtures A-E described in Table \ref{Tab:neresults}. Circles: A,  Plusses: B,  Squares: C,  
Crosses: D, Triangles: E }
\label{Fig:highv}
\end{center}
\end{figure}

\begin{figure}[h]
\begin{center}\thicklines
\epsfig{file=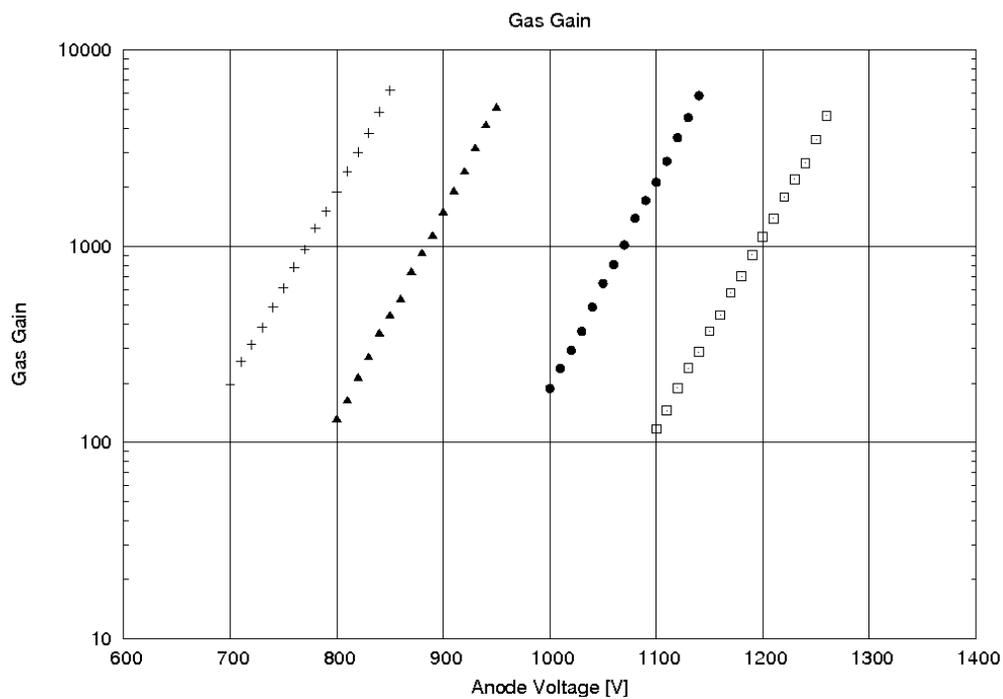,height=3.7in}
\caption{Gas Gain in mixtures containing Neon 
and
CO$_2$ and/or Nitromethane A-D described in Table \ref{Tab:gainresults}. Plusses: A,  Triangles: B,  Circles: C,  
Squares: D. }
\label{Fig:gain}
\end{center}
\end{figure}

Drift velocities for higher pressure gas mixtures of nitromethane{}-CO$_2${}-neon,
\ nitromethane{}-CO$_2${}-argon and
nitromethane{}-CH$_4$ were measured up to drift fields of
4.0 $\times$ 10$^{5}$ V/m. \ Results are shown in Figure \ref{Fig:highv}. \ The linear dependence of drift velocity on drift field persists up to
these very high fields. \ Results of linear fits for the mixtures
studied are shown in Table \ref{Tab:neresults}.  Shown in Figure \ref{Fig:gain} are gas gain curves for higher pressure mixtures of CO$_2$-neon, and neon-CO$_2$-nitromethane.  Gains as high as several thousands were obtained in all mixtures demonstrating the ability of nitromethane as a feasible, productive detector gas.

\section{Conclusion}
A negative ion drift mixture of nitromethane with CO$_2$ is
found to have linear drift velocity vs. field and to exhibit
thermal{}-limit longitudinal diffusion up to drift fields as high as
2.8 $\times$ 10$^{4}$ V/m. \ Higher pressure mixtures with argon, neon,
CH$_4$ and CO$_2$ also show linear drift
velocity up to fields as high as 4.0 $\times$ 10$^{5}$ V/m. \ All mixtures exhibit stable
operation as NITPCs.

The introduction of nitromethane as an electron capture agent will
enable the use of the NITPC as an X{}-ray polarimeter in the 1{}-10 keV
band. \ The low drift velocity of the NITPC greatly eases the
difficulty of calibrating drift velocity and substantially reduces
power consumption, which is particularly important for
satellite{}-borne astronomical instruments. \ The low diffusion of the
NITPC will also enable large{}-volume photoelectric polarimeters that
could be used, for example, as wide field{}-of{}-view instruments for
unpredictable astronomical transients, such as gamma{}-ray bursts, or
with rotation modulation collimators for high{}-resolution imaging
polarimetry of solar flares\cite{Black07}.

\bigskip

\bfseries
Acknowledgements

\textmd{This work was performed, in part, under NASA contract
}NNG07EJ03C\textmd{.}

\newpage

\newpage

\newpage

\begin{table}
\begin{tabular}{|l|l|l|l|l|l|}
\hline
Mixture & \co2 Torr & Neon Torr & Argon Torr & Methane Torr & Mobility [m$^2$/(V${\cdot}$s)]\\
\hline
\hline
A & 510 & 170 & -- & -- & 8.43 $\pm$ .01 $\times$ 10$^{-5}$ \\
\hline
B & 170 & - & 510 & -- & 9.69 $\pm$ .01 $\times$ 10$^{-5}$\\
\hline
C & 340 & 340 & -- & -- & 10.5 $\pm$ .01 $\times$ 10$^{-5}$ \\
\hline
D & 170 & 510 & -- & -- & 13.1 $\pm$ .02 $\times$ 10$^{-5}$\\
\hline
E & -- & -- & -- & 380 & 21.0 $\pm$ .07 $\times$ 10$^{-5}$\\
\hline
\end{tabular}
\caption{Fitted slopes for ion drift in mixtures of 20 
Torr
Nitromethane with \co2, Neon, Argon and Methane.}
\label{Tab:neresults}
\end{table}

\newpage

\begin{table}
\begin{tabular}{|l|l|l|l|}
\hline
Mixture & \co2 Torr & Neon Torr & Nitromethane\\
\hline
\hline
Plusses: A & 210 & 490 & -\\
\hline
Triangles: B & 170 & 510 & 20 \\
\hline
Circles: C & 340 & 340 &  20 \\
\hline
Squares: D & 510 & 170 & 20 \\
\hline

\end{tabular}
\caption{Mixture key for Figure \ref{Fig:gain}.}
\label{Tab:gainresults}
\end{table}

\end{document}